\documentstyle[11pt]{article}
   \newcommand{\be}{\begin{eqnarray}}
   \newcommand{\ee}{\end{eqnarray}}
   \renewcommand{\theequation}{\thesection.\arabic{equation}}
\begin{document}{
\begin{titlepage}
\begin{center}
{\bf {\Large Theoretical Physics Institute\\
University of Minnesota}}
\end{center}
\begin{center}
\bigskip
\begin{quote}
\raggedleft {TPI-MINN-94/6-T}
\end{quote}
\begin{quote}
\raggedleft {February 1994}
\end{quote}
\bigskip
{\bf {\Large {Instantons and fermion condensate
in adjoint $QCD_2$}}}
\end{center}
\vskip 12pt
{\centerline {\bf A.V.Smilga} }
\vskip 12pt
\parindent=20pt
\end{titlepage}

\setlength{\baselineskip}{24.2pt}
\begin{center}
Theoretical Physics Institute, University of Minnesota\\
116 Church St. S.E., Minneapolis, MN 55455, USA
\footnote{On leave of absence from ITEP, B.Cheremushkinskaya 25, Moscow,
117259, Russia.}\\

\vskip 12pt
{\bf Abstract} \\
\end{center}

 We show that $QCD_2$ with adjoint fermions involves instantons due to
nontrivial
$\pi_1[SU(N)/Z_N]~=~Z_N$. At high temperatures, quasiclassical approximation
works and the action and the form of effective  (with account
of quantum corrections) instanton solution can be evaluated. Instanton presents
a localized configuration with the size $\propto g^{-1}$. At $N=2$, it involves
exactly 2 zero fermion modes and gives rise to fermion condensate
$<\bar{\lambda}^a \lambda^a>_T$ which
falls off $\propto \exp\{-\pi^{3/2} T/g\}$ at high $T$ but remains finite.
 At low temperatures, both instanton and bosonization arguments
also exhibit the appearance of fermion condensate
$<\bar{\lambda}^a \lambda^a>_{T=0} ~\sim ~g$. For $N>2$, the situation
is paradoxical. There are
$2(N-1)$ fermion zero modes in the instanton background which implies the
absence of the condensate in the massless limit. From the other hand,
bosonization arguments suggest the appearance of the condensate for any $N$.
Possible ways to resolve this paradox (which occurs also in some 4-dim
gauge theories) are discussed.
\newpage
\section{Introduction}
Two-dimensional $QCD$ with fermions belonging to adjoint representation of
$SU(N)$ group attracted lately a considerable attention. In very interesting
recent works \cite{Kleb}, the spectrum of the theory in the large $N$ limit
has been determined. It displayed the features which are strikingly
analogous to the spectrum of 4-dimensional $QCD$ . In contrast to $QCD_2$
with fundamental fermions where the meson states lie on one Regge-like
trajectory \cite{funsp} so that
   \be
   \label{spectr}
M_n^2 ~\sim~ g^2N_c n
   \ee
and the density of states rises linearly with mass $dn/dM \sim M$, here the
number of such trajectories is infinite, and the density of states grows
exponentially with\\ mass
 \footnote{The notion of trajectories makes sense only for few first states
with small enough mass. At larger masses, the trajectories begin to overlap,
and the spectrum becomes stochastic \cite{Kleb}.}.

Of course, it is exactly the same behaviour as in large $N~~~QCD_4$ where
the number of infinitely narrow resonances also rises exponentially with
energy so that the Hagedorn phenomenon~ ---~ the appearance of limiting
temperature
above which the system cannot be heated~ --- ~ takes place. \cite{Hag}.

In this paper, we show that the adjoint $QCD_2$ bears much
resemblance with 4-dim $QCD$
describing the real world also for finite $N$. The situation is clear and
the analogy is straightforward for $N = 2$. In particular,
we show that , in contrast to what happens  in $QCD_2$ with fundamental
quarks, fermion condensate is generated here which falls down rapidly
at high $T$. The physical picture is the same as in
$QCD_4$ with only one light quark flavor \cite{Yung} and in the Schwinger
model \cite{hoso,Wipf,cond}.

The main effect leading to appearance of the fermion condensate is the
presense of instantons. They are specific for the theory with adjoint matter
and were absent in $QCD_2^{fund.}$. The topological reason for their
existence is the nontrivial $\pi_1[{\cal G}]$ where the gauge group ${\cal G}$
is $SU(N)/Z_N$ rather than just $SU(N)$ (adjoint fields are not transformed
under the action of the elements of the center), so that there are $N$
topologically nonequivalent sectors.

Instantons appear by the same token as in the Schwinger model
\cite{indus,Wipf,inst}. In the
latter, the topological reason for existence of instantons is the nontrivial
$\pi_1[U(1)]~=~Z$. The difference with nonabelian case is that, in Schwinger
model, topological charge can be written as an integral invariant
   \be
   \label{nu}
  \nu~=~ \frac{g}{4\pi}~\int d^2x~F_{\mu\nu} \epsilon_{\mu\nu}
   \ee
(it is the two-dimensional analog of the 4-dim Pontryagin class $\propto ~ \int
d^4x Tr\{F_{\mu\nu}\tilde{F}_{\mu\nu}\}$). $\nu$ is an arbitrary integer which
labels different topological sectors. In nonabelian theory, no such integral
invariant can be written ($Tr\{F_{\mu\nu}^a t^a\} ~=~ 0$). That is
understandable, of course. If such an integral invariant would exist, the
number of topologically nontrivial sectors would be infinite, but it {\em is}
finite in nonabelian case.

These new instantons which are specific for theories involving only
adjoint fields occur also in 4 dimensions. Actually, they have been known
for a long time by the nickname of 't Hooft fluxes \cite{flux}. The difference
with two dimensions is that, for $d=4$, the corresponding configurations are
not localized (they do not depend on two transverse directions), and their
action is infinite. For high T, these "planar instantons" have been studied
in \cite{bub} and also earlier in \cite{pisa} (where they were, however,
{\em misinterpreted} as real walls in Minkowsky space separating different
thermal vacua~ ---~we refer an interested reader to \cite{bub} for detailed
discussion of this question).

Topologically nontrivial fields appear in $QCD_2^{adj.}$ both
at low and at high temperatures. However, high-$T$ case is more "clean"
because quantum fluctuations are small here
, quasiclassical approximation works,  and a quantitative calculation
for the instanton contribution in the partition function is possible.

One immediate effect related to instantons is the generation of the fermion
condensate due to the presense of fermion zero modes in the instanton
background
. Recall the situation in $QCD_4$. Instantons involve there one complex zero
mode for each light fermion flavor (one for $\psi$ and one for $\bar{\psi}$).
If $N_f~=~1$, these zero modes are "absorbed" by external $\psi$--operators in
the Euclidean functional integral
   \be
   \label{Dcond}
<\bar{\psi} \psi> ~~\sim~~ \int \prod dA d\bar{\psi} d\psi ~\bar{\psi} \psi
\exp \left\{\int d^4x \left[-\frac 12 Tr (F^2_{\mu\nu}) ~+~ i\bar{\psi}
{\cal D}_\mu \gamma_\mu \psi \right] \right \}
   \ee
and we get a finite result even for very large T. If $N_f \geq 2$, there are
extra zero modes for extra flavors, and $<\bar{\psi} \psi>_{T \gg \Lambda_{QCD}
}$ is zero for massless quarks. For small $T$, ~~$<\bar{\psi} \psi>$ is
nonzero (It is the experimental fact. Theoretically, its appearance can
also be related to instanton zero modes but not in a direct way \cite{shur})
which means that the phase transition occurs.

The main observation of this paper is that the physics of $QCD_2^{adj.}$ with
$N=2$ and one
Majorana adjoint fermion flavor, is essentially the same as that of
$QCD_4$ with $N_f = 1$. High- $T$ instanton (the topologically nontrivial
configuration which minimizes the effective action) involves exactly two
zero modes
which are absorbed by external fermion operators in the functional integral
for $<\bar{\lambda}^a \lambda^a>$ and leads to exponentially suppressed
$\propto \exp\{-\pi^{3/2} T/g\}$ but nonzero result.

What happens at low temperatures ? Quantum fluctuations are large there and
only dimensional estimates can be done. Still, these estimates display
the presence of the condensate. Its value is of order $g$.  The appearance
of the condensate is also very clearly seen in the framework of bosonization
approach . It is very essential that, in contrast
to $QCD_2^{fund.}$, the bosonized version of $QCD_2^{adj.}$ does not involve
a massless field which smears away the condensate $<\bar{\psi} \psi>$ in
the former for any finite $N$.

Whereas, for $N=2$, the picture is rather clear and self-consistent, it is not
so for $N \geq 3$. High-$T$ instantons involve here $2(N-1)$ fermion zero
modes which is "larger than necessary". Similar to what happens in $QCD_4$
with $N_f \geq 2$, the extra zero modes lead to the suppression of the
condensate in the massless limit. In $QCD_4$, the statement of the absence
of the condensate at high $T$ could not be extrapolated to low temperature
region due to the presence of Goldstone bosons which display themselves in the
low temperature partition function \cite{LS}. But in $QCD_2^{adj.}$, golstones
are absent. They cannot appear in two dimensions due to the Coleman theorem
\cite{Coleman} and they do not as the generation of the fermion condensate {\em
is} not associated with spontaneous breaking of a continuous symmetry.

Assuming that {\em any} topologically nontrivial background involves exactly
$2(N-1)$ fermion zero modes and the absence of massless modes in the spectrum,
we have to conclude that the condensate is absent also at low $T$. From the
other hand, bosonization arguments display the presence of the condensate
universally for any $N$. This is a clear paradox. A possible way to resolve
it which we suggest will be discussed later in this paper.

The structure of the paper is the following. In the next section,
we fix notations and discuss the symmetries of the theory considered. In Sect.
3, the explicit form of the high-$T$ instanton for $N=2$ is obtained and the
estimate for the fermion condensate is done.
In Sect.4, we discuss the low
temperature region and show that both the instanton arguments and the
bosonization arguments imply the appearance of fermion
condensate  .
In Sect.5, we discuss characteristic field
confugurations contributing to the partition function of the theory and show
that the instantons are in some sense "confined" for strictly massless case
and are "liberated" for any small but nonzero fermion mass.
In Sect.6, we analyze the case $N \geq 3$ and display the paradox. The
paradox and
possible ways for its resolution are  discussed further in Sect.7. Conclusive
remarks are given in the last section.

\section{$QCD_2$ with real adjoint fermions}
\setcounter{equation}{0}
The lagrangian of the model reads
  \be
  \label{LQCD2}
{\cal L} = -\frac 14 F^a_{\mu\nu} F^a_{\mu\nu} + \frac i2 \bar{\lambda}^a
[ \delta^{ab} \partial_{\mu} - g\epsilon^{abc}A_{\mu}^c ] \gamma_{\mu}
\lambda^b
  \ee
where $\lambda^a_{\alpha}$ is the 2-dimensionar Majorana (real) spinor,
$\alpha~=~1,2$ ;~ $\bar{\lambda}^a~\equiv~\lambda^a\gamma^0$. It is
convenient to choose the representation
$\gamma_0~=~\sigma^2, \ \gamma_1~=~i\sigma^1$. In that case, $\gamma_5 ~=~
\gamma_0 \gamma_1 ~=~\sigma^3$\ and the left $\lambda_L ~=~ \frac 12 (1+
\gamma_5) \lambda$ and the right  $\lambda_R ~=~ \frac 12 (1 -
\gamma_5) \lambda$ components of the fermion field are described by the upper
and lower components of the spinor $\lambda_\alpha$, respectively.
The fermion part of the lagrangian can be written in terms of  $\lambda_L$
and  $\lambda_R$ as
   \be
   \label{Lferm}
{\cal L}_{ferm} = \frac i2 \left \{{\lambda}^a_L
[ \delta^{ab} \partial_{-} - g\epsilon^{abc}A_{-}^c ]
\lambda^b_L \ +\ {\lambda}^a_R
[ \delta^{ab} \partial_{+} - g\epsilon^{abc}A_{+}^c ]
\lambda^b_R \right \}
   \ee
with $\partial_\pm ~=~\partial_0 \pm \partial_1, \ A_\pm^c ~=~ A_0^c \pm A_1^c$
(Left fermions are the left movers and right fermions are the right movers).

Note (and this is very important) that, in contrast to the theory with
fundamental Dirac fermion, the lagrangian (\ref{LQCD2}) does not enjoy any
continuous global symmetry. The phase transformations $\lambda \rightarrow
\exp\{i\alpha\} \lambda$ or  $\lambda \rightarrow
\exp\{i\beta \gamma_5\} \lambda$ are not allowed as they destroy the reality
condition. The would-be currents corresponding to these transformations
$\bar{\lambda} \gamma_\mu \lambda$ and
$\bar{\lambda} \gamma_\mu \gamma_5 \lambda$ are just zero for Majorana
fermions.
One cannot also mix left and right components
$\lambda_L \equiv  \lambda_1$ and
$\lambda_R \equiv  \lambda_2$ ~---~ the lagrangian (\ref{Lferm}) is not
invariant under such a transformation.

In this respect, the situation in two dimensions differs essentially from the
4-dim case. The 4-dimensional Majorana spinor can be expressed in terms of a
complex 2-component Weyl spinor $w_\alpha$ , and the chiral phase
transformation
$w_\alpha \rightarrow e^{i\phi}w_\alpha$ is the symmetry of tree lagrangian.

There is, however, a discrete two-dimensional remnant of this 4-dim chiral
symmetry.
\footnote{I am indebted to I.Klebanov who pointed my attention to this point.}.
Either of the transformations
   \be
   \label{z2sym}
\lambda_L \rightarrow -  \lambda_L  \\ \nonumber
\lambda_R \rightarrow -  \lambda_R
   \ee
leaves the lagrangian (\ref{Lferm}) invariant. The mass term
  \be
  \label{mass}
m\bar{\lambda}\lambda ~=~ -2im\lambda_L \lambda_R
   \ee
would break this $Z_2 \otimes Z_2$ symmetry down to $Z_2$ (only {\em
simultaneous} change of sign of $\lambda_L$ and $\lambda_R$ is now allowed). We
shall see later that, even in the massless case, the symmetry
$(\lambda_L \rightarrow -  \lambda_L ,~
\lambda_R \rightarrow  \lambda_R)$ is actually not there in the full quantum
theory due to {\em anomaly} (this is true, at least, for $N=2$ theory which
we understand well).

\section{$N=2$:~Instantons and condensate at high T.}
\setcounter{equation}{0}
\subsection*{A.Preliminaries}
Let us consider the theory (\ref{LQCD2})
with two colors. As was already mentioned, the gauge-symmetry group of this
theory ${\cal G}$ is $SU(2)/Z_2~=~SO(3)$ with nontrivial $\pi_1[{\cal
G}]~=~Z_2$
. It admits therefore noncontractible topologically nontrivial field
configurations $\equiv$ instantons. All nontrivial configurations belong to one
and the same topological class. In this section, we are
interested only with high temperature case where quasiclassical description
works and quantitative estimates are possible. Euclidean path integrals are
defined on the asymmetrical 2-dim torus which is very long in spatial
direction,
$L \gg g^{-1}$, and narrow in Euclidean time direction, $\beta = 1/T \ll
g^{-1}$
{}.

To understand better how instantons appear, let us write down the high-$T$
effective potential on the {\em constant} $A_0$ - background in this theory.
The evaluation of the one-loop fermion determinant (In two dimensions, there
are no physical degrees of freedom associated with gauge fields, and the latter
do not contribute. Technically, the contribution of longitudinal degrees of
freedom $A_1^a$ cancels out the contribution of the ghosts) gives the answer
\cite{kut,kog}.
   \be
   \label{pot}
V_T^{eff}(A_0^3) ~=~ \frac {g^2}{2\pi} \left [ \left(A_0^3 + \frac{\pi T}{g}
\right )_{mod. \frac{2\pi T}{g}} - \frac{\pi T}{g} \right ]^2
   \ee
where we directed $A_0^a$ along the third isotopic axis for definiteness.
The potential (\ref{pot}) is plotted in
Fig.1. It has exactly the same form as in Schwinger model \cite{inst}
and is much analogous to the similar potential $V_T^{eff}(A_0^3) $ for pure
Yang-Mills theory in four dimensions \cite{Weiss}.

The potential (\ref{pot}) is periodic with the period $2\pi T/g$. The
periodicity is not causal. Really, the variable $A_0^a$ is canonically
conjugate to the Gauss law constraint, and the matrix
   \be
   \label{Oab}
O^{ab} ~=~ \exp \{\beta g f^{abc}A_0^c \}
   \ee
($f^{abc} \equiv \epsilon^{abc}$ for $N=2$)
may and should be thought of as the gauge transformation matrix acting on the
dynamic variables $A_1^a, \lambda^a_\alpha$. Now, the points $A^a_0 ~=~0$ and
$A_0^a ~=~\delta^{a3}~2\pi T/g$ correspond to one and the same matrix
$O^{ab} ~=~\delta^{ab}$ and are therefore physically equivalent (see Ref. \cite
{bub} for more detailed discussion).

One can consider, however, a field configuration which is $x$-dependent and
{\em interpolates} between the values $A_0^a ~=~0$ at $x ~=~-\infty$ and
$A^a_0 ~=~ \delta^{a3} ~2\pi T/g$ at $x ~=~\infty$ . It presents a
noncontractible
loop in $SO(3)$ and cannot be trivialized. Instanton is the configuration
belonging to this class with the minimal action. Usually, e.g. in 4-dim Yang
-Mills theory, the term "instanton" applies to solution of classical equations
of motion, i.e. to the configuration which minimizes the {\em tree} action.
In two
dimensions, this definition is not convenient by two reasons. First, such a
classical solution does not have nice properties ~---~it is just a constant
field strength configuration which is smeared out over the whole volume
$V ~=~\beta L$  with the very small field strength $F^3_{01} ~=~ 2\pi T/gL$
($A_0^3$ interpolates between 0 at $x~=~-L/2$ and $2\pi T/g$ at $x~=~L/2$
with constant slope). Second, quantum corrections can be taken into account
explicitly  here ~---~ at high $T$, higher loop corrections to the potential
(\ref{pot}) are small (In the exactly soluble Schwinger model, they are just
absent at any temperature). And if so, why not doing it ?
Thus, our definition of instanton is the configuration which minimizes the
{\em effective} action, quantum corrections being taken into account.

How to do that ? One may be tempted to allow the argument $A_0^3$ in Eq.(\ref
{pot}) to be $x$-dependent, add the tree-level kinetic term $\frac 12 (\partial
_x A^a_0)^2$ and solve the equations of motion for the effective lagrangian
thus obtained. Though this naive procedure
gives even the correct answer for the profile of the instanton, it cannot
be justified ~---~the expansion over derivatives of $A_0(x)$ breaks down at
the point $A^3_0 ~=~\pi T/g$ due to severe infrared singularities \cite{bub},
and the true effective lagrangian is highly nonlocal. One should proceed more
accurately.

\subsection*{B. Fermion determinant and zero modes.}
As we have seen, instanton presents a noncontractable loop $O^{ab}(x)$ in
$SO(3)$ group. In the covering $SU(2) \equiv S^3$, it corresponds to a path
which goes from the north pole $U \in SU(2) = 1$ at $x~=~-\infty$ to the
south pole $U \in SU(2) = -1$ at $x~=~\infty$. By symmetry considerations,
the path which minimizes the action should go along one of the meridians.
Each such meridian corresponds to the Ansatz
  \be
  \label{merid}
A^a_0(x) = n^aa(x), \nonumber \\
a(-\infty) ~=~0,~~~a(\infty) ~=~ 2\pi T/g,
  \ee
where $n^a$ is the unit color vector. Let us choose for definiteness
$n^a ~=~\delta^{a3}$ and calculate the fermion determinant on this
background. Minimizing the effective action thus obtained, we will find
the profile of the instanton $a(x)$ and evaluate its contribution to
the partition function.

Right from the beginning, however, we run into the problem. The matter is, the
lagrangian (\ref{LQCD2}) is well defined in Minkowski space but not in the
Euclidean space . In Euclidean space, we cannot keep the fermion fields real
{}~---~if we try to do so, the Euclidean counterpart of (\ref{Lferm}) with
$\partial_0 \rightarrow i\partial_0$ becomes complex. This problem is well
know in 4-dimensions \cite{Ramon} and its resolution is also known
\cite{VZ,LS}. One should {\em define} the integral over Majorana fields as
the square root of the determinant of the full Dirac operator. The latter is
well defined also in Euclidean space. The extraction of square root also does
not present problems here. The matter is, the spectrum of the
eigenvalue equation for the
Euclidean Dirac operator for complex adjoint fields
  \be
  \label{Direq}
\gamma_\mu^E(\partial_\mu\delta^{ab} ~-~g\epsilon^{abc}A_\mu^c)\psi_n^b(x,\tau)
{}~=~\mu_n~\psi_n^a(x,\tau)
   \ee
has a double degenerate spectrum ($\gamma_\mu^E$ are antihermitian Euclidean
$\gamma$-matrices). If $\psi_n(x,\tau)$ is a complex solution
to (\ref{Direq}), the function
   \be
   \label{degen}
\tilde{\psi}_n(x,\tau) ~=~C\psi_n^*(x,\tau)
   \ee
is also a solution with the same eigenvalue $\mu_n$. (C is the charge
conjugation matrix defined by $(\gamma_\mu^E)^*~=~-(\gamma_\mu^E)^T~=~
C\gamma_\mu^E C^{-1}$.
In 2-dimensions, $C ~=~\sigma^2$, under the particular choice
$\gamma_0^E~=~i\sigma^2, \gamma_1^E~=~i\sigma^1$.). In view of $C^*C~=~-1$, the
two solutions are linearly independent. Hence, the square root is taken without
pain:
  \be
  \label{detdef}
[ Det~\|\hat{{\cal D}}\|]^{1/2}~~=~~\left [\prod_{n}~\mu_n^2 \right ]^{1/2}
{}~~=~~\prod_{n} ~\mu_n
  \ee
where only one of the double degenerate eigenvalues $\mu_n$ is accounted for
in the product.
  Let us write the equation (\ref{Direq}) on the abelian background
(\ref{merid}). It splits apart in two:
   \be
   \label{Dirab}
\gamma_\mu^E(\partial_\mu ~-~ig\delta_{\mu 0}a)\psi_n^-
{}~=~\mu_n~\psi_n^-  \nonumber \\
\gamma_\mu^E(\partial_\mu ~+~ig\delta_{\mu 0}a)\psi_n^+
{}~=~\mu_n~\psi_n^+
   \ee
where $\psi^\pm = \psi^1 \pm i\psi^2$. [there is also the third equation for
$\psi^3$, but it is just a free one ~---~the component $\psi^3$ decouples
from the background (\ref{merid})]. It is explicitly seen that the solutions
to these two equations are related by the transformation (\ref{degen}) . The
equations are exactly the same as for 2-dim QED on the instanton background
$A_\mu(x) = \delta_{\mu 0}a(x)$ for the fermions with the charges $g$ and $-g$,
respectively.
\footnote{In the standard abelian
convention, this configuration should be called
an antiinstanton rather than instanton ~---~ its topological charge (\ref{nu})
is equal to -1. But in $QCD_2^{adj}$ with $N=2$, all noncontractible
configurations (\ref{merid}) belong to one and the same (the instanton)
topological class. To make the analogy between the nonabelian and abelian
theory more clear, we have changed the sign convention in the latter.
Also the prepotential $\phi(x)$ defined in Eq.(\ref{phi}) has
the opposite sign compared to that in Refs. \cite{Wipf,bub,inst}.}
Thus, we need not calculate the determinant anew, but rather use
the results of \cite{Wipf,inst} where the instanton Dirac determinant has
been calculated for the abelian theory:
   \be
   \label{detrel}
\left [ Det_{Ab.Ans.} \|i\hat{{\cal D}}\| \right ]^{1/2} ~=~
\left \{\left [ Det_{QED} \|i\hat{{\cal D}}\| \right ]^2 \right \}^{1/2} ~=~
 Det_{QED} \|i\hat{{\cal D}}\|
   \ee

Now, it is a proper time to note that all these determinants calculated on
the instanton background just turn to zero for strictly massless fermions
due to the presence of fermion zero modes in the spectrum ! Each of the
equations in (\ref{Dirab}) has exactly one normalizable solution with
$\mu = 0$, the left one for $\psi^-$ and the right one for $\psi^+$:
   \be
   \label{zero}
\psi_\alpha^{-(0)}(x,\tau) ~=~\left (\begin{array}{c} 1\\0 \end{array}\right )
   e^{-g\phi(x)} e^{i\pi T\tau} \nonumber \\
\psi_\alpha^{+(0)}(x,\tau) ~=~\left (\begin{array}{c} 0\\1 \end{array}\right )
e^{-g\phi(x)} e^{-i\pi T\tau}
   \ee
where
   \be
   \label{phi}
\partial\phi/\partial x ~~=~~a(x) ~-~\frac {\pi T}g
   \ee
(the $\tau$-dependence provides the correct antiperiodic boundary conditions\\
$\psi(\beta) ~=~-\psi(0)$ for the fermion fields in Euclidean time direction).
We show in the Appendix that zero mode solutions are still there also for
configurations involving small fluctuations around the abelian Ansatz
(\ref{merid}).

The presence of fermion zero modes suppresses the contribution of
topologically nontrivial sectors to the partition function exactly in the same
way as it does in $QCD_4$. To get a nontrivial result, one should introduce a
small but finite fermion mass $m \ll g$. In that case, the partition function
involves $Det \| i\hat{{\cal D}} - m\|$ rather than just  $Det \|
i\hat{{\cal D}} \|$,
and the whole result (\ref{detrel}) will be proportional to $m$.

The accurate calculation of the determinant gives the result \cite{Wipf}
   \be
   \label{detres}
\left [ Det_{Ab.Ans.} \|i\hat{{\cal D}} -m\| \right ]^{1/2} ~=~
 Det_{QED} \|i\hat{{\cal D}} -m\| \nonumber \\
\propto m\int dx~e^{-2g\phi(x)}\exp \left \{ -\frac{\beta g^2}{2\pi}
\int dy a^2(y) \right \}
   \ee
The second factor in the determinant comes from nonzero modes. In Schwinger
model, it was responsible for generating the photon mass. The first factor
is due to the zero modes. The proportionality coefficient in (\ref{detres})
can be explicitly determined (in finite box which provides infrared
regularization) if choosing a particular convention for $\phi(x)$ (the
equation (\ref{phi}) defines $\phi(x)$  only up to an arbitrary constant).
We refer the reader to Ref.\cite{Wipf} for the detailed and accurate analysis.

If we substitute now the result (\ref{detres}) in the bosonic functional
integral, calculate it, differentiate over mass and divide over the similar
functional integral for the partition function $Z_0$ in the topologically
trivial sector, we obtain the expectation value for the operator $\bar{\lambda}
\lambda$, i.e. the fermion condensate.

Let us recall how it has been done in the Schwinger model. The functional
integral in the one-instanton topological sector had the form
  \be
  \label{Z1}
Z_1 \propto Z_0 m \int \prod_y d\phi(y) \int dx~e^{-2g\phi(x)} \nonumber \\
\exp \left \{ -\frac{\beta }{2}
\int dy \phi(y) \left [ \frac {\partial^4}{\partial y^4} - \frac{g^2}\pi
\frac {\partial^2}{\partial y^2} \right ] \phi(y) \right \}
  \ee
The saddle points of this integral were determined from the equation
  \be
  \label{eqmot}
\left [ \frac {\partial^4}{\partial y^4} - \frac{g^2}\pi
\frac {\partial^2}{\partial y^2} \right ] \phi(y) ~~=~~
-2gT \delta(y-x)
  \ee
(The parameter $x$ has the meaning of the collective coordinate describing
the position of the instanton).
Substituting the solution of this equation in Eq.(\ref{phi}), we got the
result \cite{bub,inst}
   \be
   \label{inst}
a(y) ~~=~~  \left [\begin{array}{c}
\frac {\pi T}g \exp\{\mu(y-x)\},  ~~~~~~~ y \leq x \\
\frac {\pi T}g [2~-~\exp\{\mu(x-y)\}],  ~~~~~~~ y \geq x
\end{array} \right.
   \ee
where $\mu ~=~g/\sqrt{\pi}$. The function $a(y)$ is plotted in Fig.2.
The field density $E(y) = -\partial a(y)/\partial y$
is localized at $|y-x| \sim \mu^{-1}$ so that the topological charge (\ref{nu})
is equal to -1 as expected. The characteristic quantum fluctuations determined
by the integral (\ref{Z1}) are $a^{qu} \sim \sqrt{T/g}$ \cite{bub,inst} which
is much less
than the characteristic amplitude of the solution (\ref{inst})
$a^{cl} \sim T/g$ ~~so that the quasiclasical picture works.

Calculating the whole integral (\ref{Z1}) and adding the equal contribution
from the one-antiinstanton sector (in the abelian case, the relevant topology
is $\pi_1[U(1)] ~=~Z$ and instanton and antiinstanton configurations are
topologically nonequivalent), one obtains the following result for the
fermion condensate \cite{Wipf}
    \be
    \label{condab}
<\bar{\psi}\psi>_{T \gg g} ~=~ -\frac 1{\beta L Z_0} ~\frac{\partial}{\partial
m}
(Z_1 + Z_{-1}) ~=~ -2T \exp \left \{ - \frac{\pi^{3/2}T}g\right \}
    \ee
[the large factor $L$ in the denominator cancels out the large factor $L$
coming from the integration over translational zero mode of the instanton
solution (\ref{inst})]

Let us turn now to the nonabelian case. In the framework of the Ansatz
(\ref{merid}), the functional integral for $Z_1$ is basically the same as in
the Schwinger model, and its saddle point is given by the same expression
(\ref{inst}). However, two novel features appear.
First of all, besides
integrating over $\prod da(y)$ and $dx$,
we should integrate also over $dn^a$ in the Ansatz
(\ref{merid}). $n^a$ is the new collective coordinate describing the
orientation
of the instanton in color space. Naturally, rotation of $n^a$ does not
change action and corresponds to zero modes in bosonic determinant.
Let us make an estimate for the contribution
of these zero modes. The general method for such calculation is presenting
the integral over quantum fluctuations over the classical solution
(\ref{merid}), (\ref{inst}) which do not change the action
as the integral over collective coordinates $n^a$:
\cite{meas,col1,ABC}. To this end, one should express $A_0^{qu(0)}(y)$ as a
sum of two independent normalized zero modes
  \begin{equation}
  \label{Aqu0}
A_0^{qu(0)}(y) ~~=~~c_a^{(0)}~\frac{\partial A_0^{cl}(y)}{\partial n^b}
\left [ \int dy
\left ( \frac{\partial A_0^{cl}(y)}
{\partial n^c}\right ) ^2 /2 \right ] ^{-1/2}
{}~(1-n^a n^b)
   \end{equation}
and then write
   \be
   \label{mera0}
d^{(0)}A_0^{qu}(y) ~~\sim~~dc^a dc^b (1-n^a n^b) ~~\sim~~  d \vec{n}
\int dy \left ( \frac{\partial A_o^{cl}(y)}{\partial n^a} \right )^2
   \ee
The representation (\ref{merid}) is, however, not convenient for this purpose
because the zero modes $\partial A_o^{cl}/\partial n^a$ appear to be not
normalizable (this difficulty is also well known in 4-dim theories
\cite{meas}).
The paradox can be resolved by noting that the proper measure in the functional
integral is $\prod_y d^{inv.} O^{ab}(y)$ ~, $O^{ab}(y)$ being given by Eq.
(\ref{Oab}) , rather than just $\prod_{ya} dA_o^a(y)$. Thus, we have
   \be
   \label{NOab}
\int d^{(0)}O_{ab}^{qu}(y) ~~\sim~~
\int d \vec{n}
\int dy \left ( \frac{\partial O_{ab}^{cl}(y)}{\partial n^a} \right )^2
 ~~\sim \nonumber \\
\int dy \sin^2 \left [ \frac{ga^{cl}(y)}{2T} \right ] ~~\sim~~ \frac 1g
   \ee

To find the condensate, we should divide $Z_1$ by $Z_0$. The latter may be
estimated in the one loop approximation (see, however, the discussion of the
validity of this approximation later in the paper) in which case the fermion
condensate depends on the ratio of two bosonic determinants ~---~one calculated
on the background (\ref{merid}), (\ref{inst}), and the other ~---~on the
trivial
background $O^{ab}(y) = \delta^{ab}$. Thus, one should divide the result
(\ref{NOab}) by the corresponding integral in the topologically trivial
sector where only the constant harmonics of $A_0^1(y)$ and $A_0^2(y)$ should be
taken into account (the integrals over $y$-dependent parts of $A_0^1(y)$ and
$A_0^2(y)$ cancel out the contribution of {\em nonzero} modes in the bosonic
determinant in the topologically trivial sector \cite{Weiss,kut,kog,inst}).
The range of $y$ where this constant harmonic mode should be normalized is
the characteristic size of the instanton $\propto g^{-1}$ ($O^{ab} \sim
\delta^{ab}$ far away from the instanton center and the contributions of
these distances in $Z_1$ and $Z_0$ cancel out). Hence, the denominator over
which the integral (\ref{NOab}) should be divided is
   \be
   \label{NZ0}
\sim~\int_{|y-x| \sim g^{-1}} dy \left ( \frac{\partial
O^{ab}}{\partial A_0^c} \right )^2 ~\int dA_0^1~dA_0^2~ \\ \nonumber
\exp \left \{
- \frac {\beta g^2}{2\pi} \int_{|y-x| \sim g^{-1}}dy \left [
(A_0^1)^2 +(A_0^2)^2 \right ] \right \} \nonumber \\
\sim~~ \frac 1g \left (\beta g \right )^2~\frac 1{\beta g} ~\sim~\frac 1T
   \ee

Thus, rotational zero modes provide the factor $\propto T/g$ in the condensate.
In fact, this estimate could be obtained immediately using the rule of thumb
coined in \cite{col1} (see also \cite{ABC}) : each bosonic zero mode provides
the factor
$\sqrt{S_0}$ in  the measure where $S_0$ is the instanton action. In our case,
$S_0 = \pi^{3/2}T/g$ [see Eq.(\ref{condab})], and there are two rotational
zero modes. Our final result for the fermion condensate in $QCD_2^{adj.}$ with
$N=2$ at high $T$ is
  \be
   \label{condnab}
<\bar{\lambda}\lambda>_{T \gg g} ~=~ -\frac 1{\beta L Z_0}
{}~\frac{\partial}{\partial m}
(Z_1 ) ~=~ C \frac{T^2}g \exp \left \{ - \frac{\pi^{3/2}T}g\right \}
   \ee

Unfortunately, the numerical coefficient $C$ cannot be fixed here and a more
accurate calculation for the ratio of determinants which could, in principle,
be done would not help. The matter is (and this is
the second and more
serious nuisance which distinguishes the
nonabelian case compared to the exactly
soluble Schwinger model) that the partition function $Z_0$ in the
topologically trivial
sector by which the integral for $\bar{\lambda}\lambda$ should be divided
{\em can}not be determined analytically here ~---~one-loop approximation
is not justified and higher-loop effect provide a comparable contribution
in the free energy. We return to the discussion of this point in Sect.5.

  But it convenient for us to adjourn now for a while the discussion of
high-$T$ instanton physics and look first what happens in the low temperature
region.

\section{Low temperatures.}
\setcounter{equation}{0}
Consider now $QCD_2^{adj.}$ with $N=2$ at $T=0$. Let us assume that the fields
contributing to the Euclidean path integral tend to pure gauge at spatial
infinity:
   \be
   \label{pure}
i\epsilon^{abc}A_\mu^a(x)  ~\stackrel{r \rightarrow \infty}{\rightarrow}~
i\Omega^{-1}(x) \partial_\mu \Omega(x)
   \ee
with $\Omega(x) \in SO(3)$. All fields belong to one of two topological
classes:
the trivial class consisting of the fields which can be continuously deformed
to zero and the instanton class for which $\Omega(x)$ presents a
noncontractible
loop in $SO(3)$ group when $x$ goes around the large spatial circle. Another
way to look at the problem is to define the theory on large 2-dimensional
sphere. A topologically nontrivial field cannot be written as a uniform regular
expression on the whole sphere. Such a field can be described by use of
two different regular expressions defined on two patches,
the northern and the southern
hemispheres, which are glued together (related by a gauge transformation)
on the equator. The transition matrix $\Omega(\phi)$ presents then a nontrivial
loop in the $SO(3)$ group (cf. the analogous description of the Schwinger
model in Ref.\cite{indus}).

High-$T$ analysis has taught us that the fields belonging to the instanton
class
involve two fermion zero modes related to each other by the transformation
(\ref{degen})
\footnote{We return to the discussion of this point in sect.7.}.
That means that the partition function in the topologically nontrivial
sector $Z_1$ involves a factor $m$ and the fermion condensate is generated
   \be
   \label{cond0}
<\bar{\lambda}\lambda>_{T = 0} ~=~ \mp \lim_{m \rightarrow 0}\frac 1{V Z_0}
{}~\frac{\partial}{\partial m}
(Z_1 ) ~\sim~ g
   \ee
where $V$ is the volume of our 2-dim sphere.
In contrast to the high-$T$ case, one-loop calculation for $Z_1$ has no sense
here as quantum fluctuations are out of control. The estimate (\ref{cond0})
has been done purely on dimensional grounds. Two signs in Eq.(\ref{cond0})
correspond to two possible choices for the partition function:
  \be
  \label{Zpm}
Z_\pm ~=~ Z_0 \pm Z_1
   \ee
The freedom in choosing the sign is exactly analogous to the freedom of choice
of the vacuum angle $\theta$ in $QCD_4$ or in the Schwinger model.
The difference
is that here we have only two topologically distinct sectors and "vacuum angle"
can acquire onle two values: 0 or $\pi$. On the hamiltonian language, the
choices (\ref{Zpm}) correspond to two possible superselection rules imposed
on the wave functionals. The plus and minus sectors of the theory do not talk
to
each other: the matrix elements of all physical operators
between the states from different sectors are zero.

The spectrum of the theory does not include massless states, the lowest excited
state having the mass $M^{gap}$ of order $g$ \cite{Kleb}. That means that, for
large volumes $Vg^2 \gg 1$, the partition functions $Z_\pm$ enjoy the extensive
property \cite{Lee}:
  \be
  \label{ext}
Z_\pm ~\propto~\exp\{-\epsilon_\pm^{vac}(m,g) V\}
  \ee
and the finite size corrections to the vacuum energy are exponentially
small $\propto \exp\{-M^{gap}R\}$. The presence of the condensate (\ref{cond0})
implies that the function $\epsilon^{vac}(m,g)$ involves a nonzero first order
term of the Taylor expansion in $m$, and we can write for $m \ll g$:
    \be
   \label{Zexp}
 Z_\pm \propto \exp\{-m<\bar{\lambda}\lambda>_\pm V\}
   \ee
with $<\bar{\lambda}\lambda>_- = -<\bar{\lambda}\lambda>_+$, and hence

   \be
   \label{Z01}
Z_0 ~\propto~ cosh\{m<\bar{\lambda}\lambda>_+ V\} \nonumber \\
Z_1 ~\propto~ - sinh\{m<\bar{\lambda}\lambda>_+ V\}
   \ee
The result (\ref{Z01}) is the analog of the result $Z_\nu \propto I_\nu(
m|\bar{\psi}\psi| V)$ for the partition function in the sector with a given
topological charge $\nu$ for $QCD_4$ with {\em one} light fermion flavor
derived in \cite{LS}.

Note that the representations (\ref{Zexp}) and (\ref{Z01}) are valid as long
as $m \ll g~,~Vg^2 \gg 1$; the dimensionless combination
$x~=~ m|<\bar{\lambda}\lambda>_\pm| V$ may be either large or small.
The instanton
zero modes are responsible for the formation of the condensate only in the
limit when $x$ is small and
$Z_1 ~\propto~ x$. In the physical large volume
limit (large $x$), the value of the condensate
{\em is} the same but the mechanism for its formation is quite different being
related to small $\propto ~~1/|<\bar{\lambda}\lambda>_\pm| V$~~
but nonzero modes
of the Dirac operator (see \cite{LS} for detailed explanations and
discussions).

The presence of 2 fermion zero modes in the instanton background gives rise
to the 't Hooft term $\sim \bar{\lambda}^a\lambda^a \sim
\bar{\lambda}^a_L \lambda_R^a$ ~in the effective lagrangian. That means that
the $Z_2 \otimes Z_2$ symmetry (\ref{z2sym}) is in fact anomalous - quantum
corrections break it down to $Z_2$ {\em explicitly}. And that means that the
condensate $<\bar{\lambda}^a \lambda^a>$ does not break spontaneously
any symmetry
of the full quantum theory. The appearance of two sectors of the theory
(\ref{Zpm}) with opposite signs of the condensate should {\em not} be
interpreted as a spontaneous breaking because, as we have already mentioned,
these two sectors correspond to different superselection rules which
should be imposed uniformly in the whole physical space and the formation
of the "domains" separated by the "walls" so that $<\bar{\lambda}\lambda>$ is
negative to the left and positive to the right is not possible
\footnote{Note that, if the plus and minus sectors could talk to each other,
the walls between them would have a finite energy (due to the absence of
transverse directions) , and the condensate would vanish (It is not easy
to break not only a continuous but also a discrete symmetry in two
dimensions). But is does not.}.

  Again, the situation is exactly the same as in $QCD_4$ with $N_f = 1$ ~---~
the presence of the sectors with different $\theta$ in the theory should not
be interpreted as a spontaneous breaking of $U(1)$ -symmetry. $\theta$ is
one and the same in the whole physical space and the spatial fluctuations of
$\theta$ (which would give rise to Goldstone bosons) are not possible.

  The existence of the condensate is also clearly seen in the framework of
bosonization approach. Since \cite{Witten}, it is known that a theory involving
Majorana fermion fields $\lambda^a$ is dual to some other theory involving
the bosonic field presenting an orthogonal matrix $\Phi^{ab}$. The correlators
of all fermion bilinears in the original theory coincide identically with the
correlators of their bosonized counterparts in the bosonized theory. For the
scalar bilinear $\bar{\lambda}^a \lambda^b$, the correspondence rule is just
  \be
  \label{corra}
\bar{\lambda}^a \lambda^b \equiv \mu \Phi^{ab}
  \ee
where $\mu$ depends on the normalization procedure for the operator
$\Phi^{ab}$.
$\mu$ is of order $g$ ~if the normalization convention $<\Phi^{ab}>_{vac}
 = \delta^{ab}$
is chosen. It is obvious then that
  \be
  \label{condb}
<\bar{\lambda}^a \lambda^a>_{vac} ~\sim~\mu~\sim~g
  \ee

  Note the difference with the theory involving fundamental Dirac fermions.
For $QCD_2^{fund}$, the bosonization rule is not (\ref{corra}) but rather
   \be
   \label{corrf}
\bar{\psi^i} \psi^j ~\equiv~ \mu U^{ij} \exp \left (i \sqrt{\frac{4\pi}N} \phi
\right )
   \ee
where $U$ is the unitary $SU(N)$ matrix, and $\phi$ is a light color singlet.
In that case, the normalization mass $\mu$ is not $g$ but depends on
the mass of the
scalar singlet which in turn depends on the fermion mass $m$.
Both $\mu$ and the light singlet mass tend
 to zero in the limit $m \rightarrow 0$ for any
finite $N$ (and the singlet becomes sterile) \cite{Frish}. The condensate
$<\bar{\psi}^i \psi^i>_{vac}$ also tends to zero in the massless limit.
One can say that the light singlet $\phi$ smears the condensate away
\footnote{The absence of the condensate in $QCD_2^{fund}$ is, of course,
natural. The condensate would break spontaneously the global chiral symmetry,
and such a breaking is not allowed in two dimensions \cite{Coleman}}.

But in the adjoint theory, all fields in the spectrum are massive and the
condensate (\ref{condb}) survives.

The rapid fall-off of the condensate at high temperature as given by Eq.(\ref
{condnab}) is also naturally explained in the bosonization language. Taking
into account finite $T$ effects ~---~namely, the presence of excited states in
the heat bath, the thermal average $<\Phi^{ab}>_T$ is no longer $\delta^{ab}$,
but can acquire any value on the $SO(3)$ group with almost equal (at high
$T \gg g$) probability, and
  \be
   \label{cbhT}
<\Phi^{aa}>_{T \rightarrow \infty} ~\rightarrow~ \int d^{inv} \Phi~
\chi^{adj}(\Phi) ~=~0
   \ee
As follows from Eq.(\ref{condnab}), for high but finite $T$ the direction
$\delta^{ab}$ in the group is still a little bit preferred, and the condensate
is still nonzero though exponentially small. The physical picture is exactly
the same as in the Schwinger model where the quantitative calculation is
possible at any temperature \cite{cond}.

\section{High-$T$ partition function.}
\setcounter{equation}{0}
All the arguments of the previous section which have led to the results
(\ref{Zexp}) and (\ref{Z01}) can be repeated without change also for high
temperatures. We only have to substitute $\beta L$ for $V$ and
$<\bar{\lambda}\lambda>_T$ for $<\bar{\lambda}\lambda>_{vac}$.
Let us look how the
partition functions (\ref{Z01}) behave when the spatial volume $L$ is very
large,
  \be
  \label{xT}
x_T ~=~ m\beta L |<\bar{\lambda}\lambda>_T| ~\gg~ 1
  \ee
The $cosh$ and $sinh$ functions in Eq.(\ref{Z01}) can be expanded in the series
and, if $x_T$ is large, the number of the terms in the series to be taken into
account is also large. Each such term is
  \be
  \label{term}
Z^{(k)} ~=~ \frac{(-m\beta L <\bar{\lambda}\lambda>_T)^k}{k!}
  \ee
where $k$ is even for $Z_0$ and odd for $Z_1$. The series converge at $k \sim
x_T$. The contribution (\ref{term}) in the partition function can be
interpreted as being due to $k$ instantons (\ref{merid}).
Each instanton brings about the factor $m$ from the fermion zero mode and
the factor $L$ from the translational bosonic zero mode in the partition
function.
The instantons are
very well
spatially separated, the characteristic inter-instanton distance being of
order $L/k^{char} ~\sim~ 1/(m\beta |<\bar{\lambda}\lambda>_T|) ~\gg~g^{-1}$.
 Thus, we see that
the characteristic field configurations in the high-$T$ partition function
present a rarefied noninteracting instanton gas . Naturally, the total
number of instantons is even for $Z_0$ (the configuration is topologically
trivial) and odd for $Z_1$.

The same picture is valid in high-$T$ Schwinger model \cite{inst} and in
high-$T$ $QCD_4$ \cite{shur}. Note that we cannot extrapolate it to low
temperatures. When $T < g$, the characteristic separation between instantons
is of the same order as their size $\sim g^{-1}$, and their interaction (as
well as distortion of their form due to quantum fluctuations) cannot be
neglected. Instead of a rarefied instanton gas, we have a dense strongly
interacting instanton liquid \cite{shur,vort}.

It is interesting to look also at the limit when $L$ is kept large but finite
and $m$ tends to zero. In strictly massless theory, $Z_1$ vanishes and $Z_0$
has no trace of instantons at all. The explanation is simple. Consider the
contribution of two well separated
instantons to $Z_0$. The zero modes of individual instantons are now shifted
from zero, but the shift is tiny:
   \be
   \label{shift}
\mu^{quasizero}(R) ~\sim~ \exp\{-\pi TR\}
   \ee
where $R$ is the inter-instanton separation. Thus, the large $R$ configurations
provide exponentially small contribution to the path integral, instantons
are "confined" and cannot be separated from each other. (The same phenomenon
occurs in the Schwinger model \cite{bub,inst}. For $QCD_2^{adj}$, it has been
actually observed in Ref.\cite{kog}).

If $m \neq 0$, the contribution of the two-instanton contribution to the path
integral ceases to depend on $R$ as soon as $\mu^{quasizero}(R) \ll m$. If $m$
is large enough (the condition (\ref{xT}) is fulfilled), two-instanton
contribution dominates over zero-instanton one: instantons are "liberated".

Now, the time has come to pay our old debt and to discuss nonperturbative
effects in $Z_0$ for the massless theory (one can forget about instantons
till the end of this section). Let us estimate free energy density
$F = -TL^{-1} \ln Z_0 $ of the theory at high temperature. In the leading
order, it is given just by the free fermion loop and is of order $T^2$. But
what are preasymptotic effects ? It is instructive to consider first Schwinger
model. In the bosonized language, it is just a theory of free scalars with
the mass $\mu = g/\sqrt{\pi}$. At finite $T$, they are excited and the exact
expression for $F$ is
   \be
   \label{free}
F_{SM}(T) ~=~ \int_{-\infty}^\infty \frac{dp}{2\pi}\ln\left [ 1 -
e^{-\beta \sqrt{p^2+\mu^2}} \right ]
   \ee
Its high-$T$ asymptotics is
   \be
    \label{freehT}
F_{SM}(T \gg \mu) ~=~ - \frac{\pi T^2}6 \left [ 1 ~-~
\frac{3 \mu}{\pi T} ~+~ o \left (
\frac \mu T \right ) \right ]
   \ee
When $T \sim \mu$, subleading effects are essential.

For $QCD_2^{adj}$, the qualitative estimate is the same, but we cannot
determine now the coefficient of preasymptotic term exactly: the mass of
bosons in the spectrum cannot be determined analytically, and their interaction
cannot be neglected. Thus, we can only write
   \be
   \label{FQCD}
\Delta F_{nonpert}^{adj} (T) ~\sim ~ gT
   \ee
That is the same
uncertainty which prevented us to determine the exact coefficient in
Eq.(\ref{condnab}): the uncertainty in $Z_0$ is
   \be
   \label{unc}
\sim \exp\{-\beta \Delta F_{nonpert}(T)~g^{-1} \} ~\sim~ 1
   \ee
($g^{-1}$ is the instanton size where the background field (\ref{merid})
differs essentially from zero and the determinants of fluctuations in $Z_1$
and $Z_0$ are different).
\footnote{Note that uncertainties of essentially the same kind in the
determination of the instanton measure appear also in $QCD_4$ {\em when the
size of the instanton $\rho$ becomes comparable with the characteristic
scale of the theory}. The corrections to the measure are of order
$\rho^4 \epsilon^{vac}_{QCD} \sim \rho^4 \Lambda_{QCD}^4$ \cite{vain}.
When $\rho \Lambda_{QCD} \sim 1$, the situation is out of control.}

\section{$N~\geq~3$.}
\setcounter{equation}{0}
\subsection{A. High T.}
Let us repeat the analysis of Sect.3 for higher color groups.
Consider first the
case $N=3$. The effective potential on the constant $A_0$ background has been
calculated in Ref.\cite{kut}. For $N=3$, it depends on 2 group invariants:
$A_0^aA_0^a$ and $d^{abc}d^{ade} A_0^b A_0^c A_0^d A_0^e$. It is convenient to
choose the matrix $A_0^a t^a$ in the diagonal form
  \be
  \label{diag}
A_0^a t^a ~~=~~diag~(a_1,~a_2,~a_3), ~~~~~~~~~~~~~~~~ \sum_i a_i ~=~0
  \ee
and write the effective potential as a function of $a_i$ (or, if you will,
as a function of $A_0^3$ and $A_0^8$). The result is
   \be
   \label{pot3}
V(a_i) ~~=~~\frac{g^2}{2\pi}~\sum_{i>j}^3 ~\left [\left (
a_i~-~a_j~+~\frac{\pi T}g
\right )_{mod. \frac{2\pi T}g}~-~\frac{\pi T}g \right ]^2
   \ee
This potential has a hexagonal symmetry. The structure of its minima is shown
in Fig.3.

What is the proper range of integration over $A_0^a$ in the functional integral
? As was discussed earlier, the proper integration variable is not $A_0^a$ but
rather the adjoint gauge transformation matrix (\ref{Oab}) (which was the
orthogonal matrix in the case $N=2$). To count each such matrix only once, we
should restrict the range of integration by the "small"
Weyl cell (marked out by the dashed lines inside the solid triangle in Fig.3)
which is spread out over the whole $SU(3)/Z_3$ group by the transformations
from the torus of the group with nonzero $A_0^{1,2,4,5,6,7}$.

Note that in the general case where the theory involves also fundamental
matter fields, the integration goes over unitary matrices
$U~=~\exp\{i\beta g A_0^a t^a\}$ which are different in the three different
classes of minima:
  \be
  \label{Umin}
U_0 ~=~1,~~~U_\Box~=~e^{2\pi i/3},~~~U_\triangle~=~e^{-2\pi i/3}
  \ee
In a theory with fundamental matter, these three sets of points mark out
physically
different gauge transformations, the counterpart of Eq.(\ref{pot3}) would
also be different at these points: $V^{fund.}(U_0) ~\neq~  V^{fund.}(U_\Box)
{}~\neq~ V^{fund.}(U_\triangle)$, and the proper integration region would be
the
standard Weyl cell (solid triangle in Fig.3) + transformations from the
torus.

But in $QCD_2^{adj.}$, the proper gauge group is $SU(3)/Z_3$ rather than
$SU(3)$, and all minima of the potential (\ref{pot3}) [which occur at the
points (\ref{Umin})] should be identified.
There are, however, noncontractible Euclidean configurations which interpolate
between different center elements (\ref{Umin}) of the unitary group so that,
say,
   \be
   \label{inter}
U(x = -\infty) ~=~ 1,~~~~~~~~~U(x = \infty) ~=~e^{2\pi i/3}
   \ee
The configuration (\ref{inter}) presents a nontrivial loop in the $SU(3)/Z_3$
{}~-~space. For $N=3$, there are two different topologically nontrivial
classes:
the configurations (\ref{inter}) which may be called instantons and the
configurations interpolating between 1 and $e^{-2\pi i/3}$ which may be called
antiinstantons (double instanton configurations are topologically equivalent
to antiinstantons).

Consider a representative of the instanton class which has the form
   \be
   \label{inst3}
t^a A_0^a(x) ~~=~~ \frac 13 a(x)~ diag(1,1,-2) \nonumber \\
a(-\infty) ~=~ 0,~~~~a(\infty) ~=~ \frac{2\pi T}g
   \ee
It corresponds to going upwards along the vertical side of the solid triangle
in Fig.3 with the transformation from the torus being fixed to be trivial
(so that different points on the side correspond to all different elements
of the group $SU(3)/Z_3$)

Let us estimate the fermion determinant in this background field. The
eigenvalue equation for the Euclidean Dirac operator [the analog of
(\ref{Dirab})] on the background (\ref{inst3}) has the form
   \be
   \label{Dirab3}
\gamma_\mu^E(\partial_\mu ~\pm~iga(x)\delta_{\mu 0})\psi_n^{4\pm i5}
{}~=~\mu_n~\psi_n^{4\pm i5}  \nonumber \\
\gamma_\mu^E(\partial_\mu ~\pm~iga(x)\delta_{\mu 0})\psi_n^{6\pm i7}
{}~=~\mu_n~\psi_n^{6\pm i7}
   \ee
and the components $\psi^{1,2,3,8}$ decouple from the background.

We see that the Dirac equation admits now not one but {\em two} pairs of zero
modes (\ref{zero}). That means that the partition function in the instanton
sector involves now the factor $m^2$ rather than just $m$. And that means that
the contribution of topologically nontrivial sectors in the condensate is
  \be
  \label{cond3}
<\bar{\lambda}^a \lambda^a>_{T \gg g}^{N=3} ~=~ -\frac 1{\beta L Z_0}
{}~\frac{\partial}{\partial m}
(Z_I ~+~Z_{A}) ~\propto~ m
  \ee
and turns to zero in the massless limit. The situation looks the same as in
$QED_2$ with two Dirac charged fermions where the fermion condensate is
zero by the same reason.

Similar analysis can be done also for larger $N$. The generalization of the
Ansatzes (\ref{merid}), (\ref{inst3}) for any $N$ is
   \be
   \label{instN}
t^a A_0^a(x) ~~=~~ \frac 1N a(x)~ diag(1,1,\ldots,1-N), \nonumber \\
a(-\infty) ~=~ 0,~~~~a(\infty) ~=~ \frac{2\pi T}g
   \ee
which supports $N-1$ pairs of fermion zero modes. The determinant has the same
structure as in the Schwinger model with $N-1$ flavors, and the contribution
to the condensate is
  \be
  \label{condN}
<\bar{\lambda}^a \lambda^a>_{T \gg g}^{N} ~\propto~ m^{N-2}
  \ee
which vanishes in the massless limit.
\footnote{For $N > 3$, the leading contribution to the condensate comes not
from instantons but just from the topologically trivial sector. The latter
gives $<\bar{\lambda} \lambda> ~\propto~ m$ for any $N$ (cf. Eq.(8.22) in
Ref.\cite{LS})}

Thus, at high temperatures, fermion condensate seems not to be
formed in $QCD_2^{adj.}$ with $N \geq 3$.

\subsection{Low T: The paradox.}
The bosonization arguments of Sect.4 which have led to the conclusion of
existence of the fermion condensate for $N=2$ can be transferred without
essential change to larger $N$. The theory involves now the set of $N^2-1$
Majorana fermion fields. Staying in the framework of
the original Witten's paper \cite{Witten}
where only free fermions were discussed, we would have to put
such a set of field
into correspondence to the boson fields presenting orthogonal $SO(N^2-1)$
matrices . In the case when the fermions interact with gauge fields, it is
more convenient, however, to write the bosonized theory in terms of the fields
   \be
   \label{Fiab}
\Phi^{ab} = Tr\{t^a U t^b U^+\}
   \ee
where $U \in SU(N)$ and $\Phi \in SU(N)/Z_N$. This modified bosonization
procedure has been worked out in \cite{Brown}. $\Phi^{ab}$ is dual to the
scalar bilinear $\bar{\lambda}^a \lambda^b$ as written in Eq.(\ref{corra}). As
earlier, $\mu \sim g$ and the estimate (\ref{condb}) for the fermion condensate
is valid.

Again, the spectrum of the theory involves a gap and, in the limit $m \ll g,
Vg^2 \gg 1$, the partition function can be written in the form
   \be
   \label{Z}
Z \propto \exp\{-m <\bar{\lambda} \lambda> V \}
   \ee
both for large and for small values of ~$m |<\bar{\lambda} \lambda>| V$.

But that contradicts instanton arguments.

Let us consider for simplicity the case $N=3$. There are 3 topological classes:
the trivial, the instanton and the antiinstanton. In the topologically
trivial sector, the partition function
  \be
  \label{Zmod}
Z_0 ~\propto~ \left < \prod_n (\lambda_n^2 + m^2) \right >,~~~~\lambda_n \neq 0
  \ee
is expanded in the even powers of $m$. The expansion of $Z_I$ and $Z_A$ in $m$
also starts from the term $\propto m^2$ due to the presence of
2 pairs of fermion zero modes. It is absolutely not clear how the linear
term in the expansion of the exponential (\ref{Z}) can appear.

Thus, bosonization arguments tell that the condensate {\em is} formed whereas
the instanton arguments tell that it {\em is} not formed.

\section{Confronting the controversy.}
\setcounter{equation}{0}
 The paradox appeared when putting together the following premises:
\begin{enumerate}
\item Validity of topological classification.
\item The presence of $2(N-1)$ zero modes in the instanton sector.
\item Bosonization arguments displaying the presence of condensate.
\item Absence of massless states which allowed us to write the partition
function in the extensive form (\ref{Z}) also for small values of exponent.
\end{enumerate}

The only way to resolve the paradox is to invalidate one of these premises.

For example, in the conventional $QCD_4$ with several flavors where instantons
involve $N_f$ zero modes, their contribution to the partition function is
$\propto m^{N_f}$ but the condensate is still generated without any paradox
because the premise 4 is false. There are Goldstone states in the spectrum
which lead to finite volume effects which are essential in the region of
small ~$m |<\bar{\psi}\psi> | V$ ~and the partition function cannot be written
in the extensive form (\ref{Z}) but has a more complicated
srtucture \cite{LS}.
But in our case, no continuous symmetry is broken spontaneously and there are
no goldstones.

At first sight, the weakest point is the second premise. We have obtained
$2(N-1)$ zero modes by solving explicitly the Dirac equation in a particular
background. We have also checked that the zero modes are stable with respect
to small deformations of background (see Appendix). But we cannot write down
an index theorem which would enforce the presence of $2(N-1)$ zero modes for
{\em any} background belonging to the instanton class. The "normal" index
$n_L^0 - n_R^0 \propto \int Tr\{F_{\mu \nu}^a t^a \} \epsilon_{\mu \nu} d^2x$
is just zero in $QCD_2$ [indeed, we have established the presence of $N-1$
left-handed and $N-1$ right-handed zero modes related to each other by the
transformation (\ref{degen}) ], and we do not know of any other relevant
integral invariant.

Thus, we cannot rule out that, for some fields belonging to the instanton class
and located at some finite distance from the abelian Ansatz in the Hilbert
space, the number of zero modes is less which would allow the generation of
the condensate.

We think, however, that it is not the case, and there {\em is} some index
theorem prescribing the existence of exactly $2(N-1)$ zero modes, only we are
not clever enough to unravel it. The reason why we  believe it is the
following .

The paradox discovered is actually not specific for $QCD_2^{adj}$. The same
paradox appears also in some 4-dim gauge theories where the conventional
Atiah-Singer theorem works and the analog of our premise 2 is certainly valid.

As we have already mentioned, there is no paradox in the conventional $QCD$.
Consider, however, supersymmetric $d = 4$,
$N = 1$ nonabelian Yang-Mills theories
involving a Majorana fermion in the adjoint representation of the gauge
group. The paradox {\em does} not arise when the group is unitary. Let us
understand why.

At first sight, it does. The fields belonging to the instanton class involve
$2N_c$ fermion zero modes (the trace $Tr\{T^aT^a\}$ which enters the index
theorem differs, for the generators $T^a$ in the adjoint representation,
by the factor $2N_c$ from the analogous trace for the fundamental
representation). That means that the instanton contribution to the partition
function involves a factor $m^{N_c}$. From the other hand, exact
supersymmetric Ward identities tell us that the correlator $<\bar{\lambda}
\lambda(x_1) \ldots \bar{\lambda}\lambda(x_{N_c}> $ does not depend on $x_i$.
The
computation in the instanton background gives nonzero result which implies that
the correlator does not vanish also when all $|x_i - x_j|$ tend to $\infty$
\cite {sinst}. And that implies the presence of the condensate
$<\bar{\lambda}\lambda>$ . As it does not break spontaneously any exact
symmetry
of the quantum theory, no massless states appear, the extensive representation
(\ref{Z}) for the partition function is valid, and we cannot reproduce the
linear in mass term in the expansion of Z
when taking into consideration only the fields with integer winding number.

The paradox is resolved in this case by noting that, for a theory involving
only adjoint fields, the fields carrying {\em fractional} winding
numbers  $\nu = \pm1/N_c, \pm2/N_c, \ldots$ are equally admissible \cite
{toron,Cohen,Zhit,LS}. The reason is, again, that the gauge group here
is actually $SU(N)/Z_N$ rather than $SU(N)$ and the gauge transformation
matrices
differing by an element of the center are undistinguishable. The configurations
with $\nu = \pm1/N_c$ involve only 2 fermion zero modes and {\em are}
responsible for the formation of fermion condensate for small
$m|<\bar{\lambda}\lambda>|V$.

The situation is much worse, however, for higher orthogonal and exceptional
groups. The simplest example where the problem appears is the SYM theory with
$SO(7)$ gauge group \cite{O7}. The instantons involve here 7-2 = 5 pairs of
zero
modes and the corresponding contribution to the partition function is
$\propto m^5$. The group $SO(7)$ does not have a nontrivial center and, in
contrast to what we had for $SU(N)$ groups, we cannot pinpoint a topological
field configuration with winding number $\nu = 1/5$. Things are not better
when $N > 7$. Thus, $SO(N \geq 7)$
4-dim SYM theories are as paradoxical as $QCD_2^{adj}$ with $N \geq 3$.

We present here another very simple 4-dimensional example where the paradox
also appears. Consider the $SU(2)$ Yang-Mills theory involving a Dirac fermion
$\psi$ belonging to the color representation with isospin $I = 3/2$. Suppose
that  the fermion condensate $<\bar{\psi} \psi> $ is formed here. Like in
the conventional $QCD_4$ with $N_f = 1$, it breaks only $U_A(1)$ subgroup
of the chiral symmetry group which is anyway anomalous, and no massless
states appear. From the other hand, comparing $Tr\{T^aT^a\} = I(I+1)(2I+1)$
in the representation with $I = 3/2$ with the same trace for $I = 1/2$, we
see that the instantons involve here 10 fermion zero modes and provide the
contribution $\propto m^{10}$ to the partition function. There is no way
to get the fermion condensate in the path integral framework.

Of course, the paradox here is not so prominent as in two other theories
considered above. It appeared when {\em assuming} that the condensate is
generated. The assumption looks natural ~---~the dynamics of the theory is
rather similar to that of conventional $QCD$ with $N_f = 1$ where the
condensate
is formed, but there is also a distinction. The first coefficient in the
Gell-Mann-Low function
  \be
  \label{GML}
b ~=~ \frac {22}3 - \frac 23 \times 10 ~=~ \frac 23
  \ee
is comparatively small here (though the theory is still asymptotically free)
which may after all prevent the formation of fermion condensate. And, in
contrast to two previous cases, we cannot present solid independent theoretical
arguments that the condensate is formed. Thus, this theory may serve only as an
additional indication that something is grossly wrong in our understanding;
we could not claim that solely on its basis.

But, for $SO(N \geq 7)~ SYM$ and for $QCD_2^{adj}$ with $N \geq 3$, the
situation is really misterious.

We cannot say that we understand how this mistery is resolved. But , if there
is a universal reason which resolves it in both theories, the only one we
can think of is that the premise 1 in the list in the beginning of this section
is false. Perhaps, there are some singular field configurations which
contribute
to the path integral and which cannot be classified by topological
considerations. If these unspecified configurations have only one pair of
fermion zero modes, the condensate may be generated.
One argument in favor of this guess comes from the observation
that, in strong coupling theory, fields fluctuate wildly  and the topological
classification which is based on the assumption that the fields are smooth
and regular may be not true.

Suggestions that this may happen can be found in the literature. In particular,
Crewther \cite{crew} and Zhitnitsky \cite{Zhit} argued that, for the
conventional $QCD_4$ with $N_f$ light flavors with equal mass, field
configurations carrying winding number $1/N_f$ (obviously, such fields cannot
be described in topological terms) can be relevant. Actually, we do not see
compelling reasons to assume this for standard $QCD$ ~---~ the usual
description including only the fields with integer winding numbers works
perfectly well there. But for $QCD_2^{adj} $ with $N \geq 3$, for $SO(7) $
4-dim SYM , and may be for $SU(2)$ 4-dim gauge theory with Dirac fermions
belonging to the representation $I = 3/2$ of the color group, we are kind
of forced to think in this direction. What is absolutely unclear by now is
in what respects path integral dynamics of these paradoxical theories
differs from that in standard QCD and other well-studied theories where no need
of invoking  exotic nontopological fields arises.

\section{Conclusions.}
\setcounter{equation}{0}
The $SO(3)~ ~~QCD_2^{adj}$ which we analyzed first in this paper presents no
problems. The picture is self consistent: the instantons which are present
there due to nontrivial $\pi_1[SO(3)] = Z_2$ involve two fermion zero modes
and lead to the formation of the fermion condensate. This condensate falls
down as the temperature increases [see Eq.(\ref{condnab})] but never turns to
zero. Qualitatively, the same follows from bosonization arguments. This
model can serve as a remarkably good playground which may allow us to
understand better the physics of $QCD$ (in particular, of $QCD$ with only one
quark flavor). For example, lattice simulations of this theory would be
very interesting. One could try to calculate the fermion condensate on the
lattice at zero and at high temperature and compare the numerical results
with theoretical prediction (\ref{condnab}). Such simulations are {\em much}
simpler than in 4 dimensions and could provide an independent test for the
whole lattice technology.

For $N \geq 3$, we encountered an explicit paradox: the existence of the
condensate follows from bosonization arguments but we could not get it in
the path integral approach. As was discussed in details in sect.7 of this
paper, a similar paradox displays itself also in some 4-dimensional gauge
theories. Its satisfactory resolution could bring about a progress
in our understanding of quantum field theory in general.

In conclusion, we note that , if we would believe in the bosonization
arguments at low temperatures and in the instanton arguments at high
temperatures (at high $T$, quasiclassical approximation works and one could
think that it still suffices to consider only smooth topological
field configurations), the conclusion of the existence of the {\em phase
transition} in the theories with $N \geq 3$ would follow~---~ at some
temperature $T_c$, the condensate would vanish and stay zero beyond it. But,
at the present level of understanding, we cannot really claim it is true.

If nontopological fields contribute to the path integral also at high
temperatures, there is no phase transition but only a crossover where the
condensate falls down but never turns to zero (as it {\em is} the case for
$N =2 $). As $N$ grows, the crossover is expected to become more and more
sharp. Its temperature is estimated as
   \be
   \label{cross}
T^* ~~\sim~~g \sqrt{N}
   \ee
(a natural mass scale of the theory). In the limit $N \rightarrow \infty,~~
T^* \rightarrow T_H$, the Hagedorn limiting temperature.

 \section{Acknowledgements.}
I am indebted to I.Klebanov and I.Kogan for illuminating discussions on the
early stage of this work and to M.A.Shifman for the discussion of SYM with
orthogonal groups. It is a pleasure for me to thank TPI department of the
University of Minnesota where this work has been done for warm hospitality.

\setcounter{section}{0}
\setcounter{equation}{0}
\renewcommand{\theequation}{A.\arabic{equation}}
\section*{Appendix.}
We want to show here that the zero modes (\ref{zero}) and their conterparts
for larger $N$ are stable
under small deformations of the abelian high-$T$ instanton background
(\ref{merid}), (\ref{instN}).
Consider first the case $N = 2$. Choose as earlier $n^a = \delta^{a3}$ in Eq.
(\ref{merid}) and deform it in the transverse direction in the color space so
that
   \be
   \label{deform}
A_0^a(x) = \delta^{a3} a(x) + (1 - \delta^{a3}) b^a(x)
   \ee
with $b^a(-\infty) = b^a(\infty) = 0$ and $b \ll a$ for all $x$. Then the
deformation $b^a(x)$ has no projection on the global gauge rotation modes
discussed at length in sect.3.

For $b^a(x) = 0$, the Dirac eigenvalue equation (\ref{Dirab}) had two zero
mode solutions (\ref{zero}). With $b \neq 0$, the solutions are modified.
Unfortunately, in contrast to the more simple abelian case \cite{indus,Wipf},
we cannot solve the zero mode equation explicitly for any gauge field
background.
What we can do is to develop a perturbation theory in the small parameter
$b/a$ and find the solution as the series in this parameter:
   \be
   \label{expan}
\psi^{a(zero)} ~=~ \psi^{a(zero)}_0 ~+~
\psi^{a(zero)}_1~+~ \psi^{a(zero)}_2 ~+~\ldots
   \ee

Let us start, for definiteness, from the solution
$\psi^{-(zero)}_0(x,\tau)$ and find the corresponding
$\psi^{a(zero)}_1(x,\tau)$. It satisfies the equation
   \be
   \label{Dirac1}
\left [( \partial_0 \sigma_2 + \partial_x\sigma_1) \delta^{ab} ~-~
g\epsilon^{ab3} a(x)\sigma_2 \right ]
\psi^{b(zero)}_1(x,\tau) \nonumber \\
{}~=~-i \frac g2 \delta^{a3} b^+(x) \sigma_2
\psi^{-(zero)}_0(x,\tau)
   \ee
We see that only the component $\psi^{3(zero)}_1$ appears. It is left-handed
as $\psi^{-(zero)}_0$ was and also has the same $\tau$-dependence $\propto \exp
(i\pi T \tau)$. The solution of
(\ref{Dirac1}) is
   \be
   \label{psi1}
\psi^{3(zero)}_1(x,\tau) ~=~ - \frac g2 e^{\pi T x} \int_x^\infty b^+(y)
\psi^{-(zero)}_0(y,\tau) e^{-\pi T y} dy
   \ee
It is easy to see that $\psi^{3(zero)}_1$ has the same asymptotics
$\propto \exp\{-\pi T|x|\}$ at $|x| \rightarrow \infty$ and is normalizable.

Generally, the n-th term of the series (\ref{expan}) $\psi^{a(zero)}_n(x)$ is
related to $\psi^{a(zero)}_{n-1}(x)$
by a similar integral kernel which provides
the asymptotics $\propto \exp\{-\pi T|x|\}$ for $\psi_{n}$ if $\psi_{n-1}$ had
such, and the normalizability of the deformed zero mode is proven by induction.

For larger $N$, the analysis is quite similar. The integral kernels are a
little
bit different for different $\psi_0^{a(zero)}$~---~ the different
color components
of the deformation $b^a(x)$ enter, but the result is the same:
if the perturbation is small,
all $2(N-1)$ different zero modes remain normalizable and are there in the
spectrum.

Certainly, this analysis cannot rule out bifurcations in the space of zero
modes when the perturbation is large enough so that the number of zero modes
would be less than $2(N-1)$ for some $b$, but we do not think
that this possibility
is realized (see the main text for more detailed discussion).

\newpage
\section*{Figure captions.}

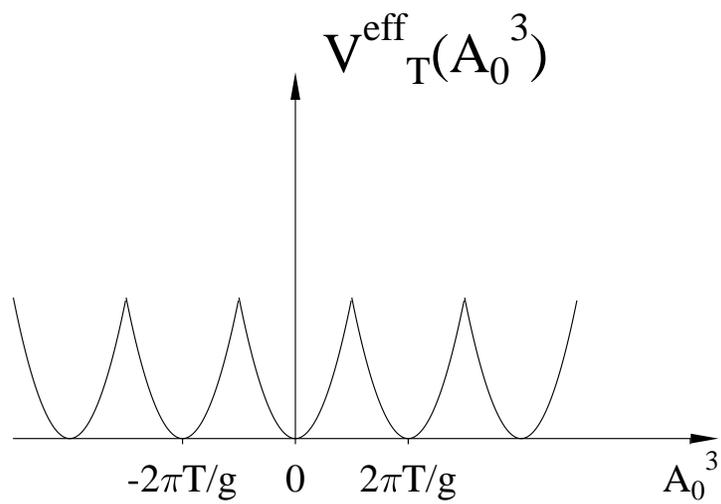
{\bf Fig.1.} Effective potential in adjoint $SU(2)$ theory at high $T$.

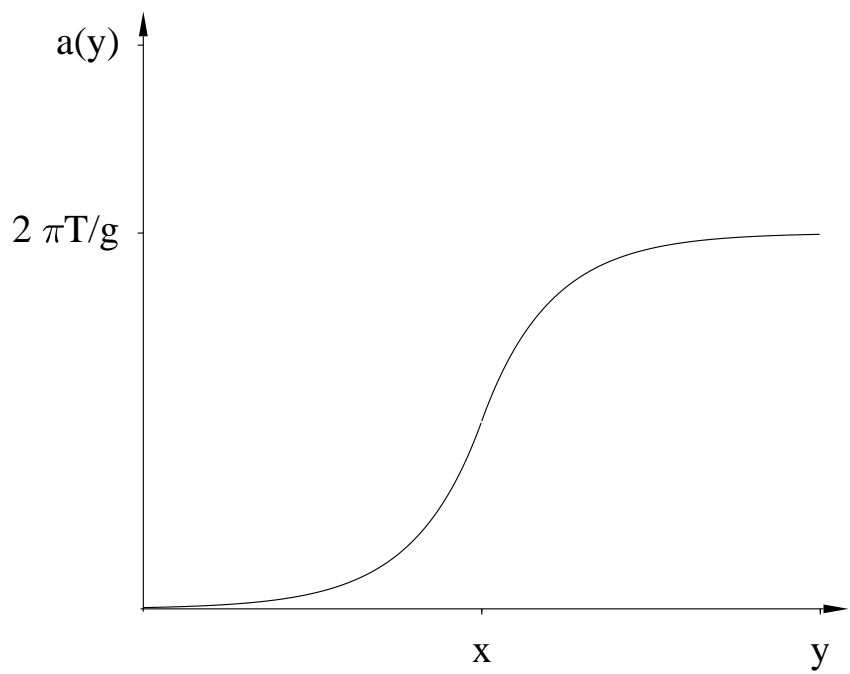
{\bf Fig.2.} High-$T$ instanton.

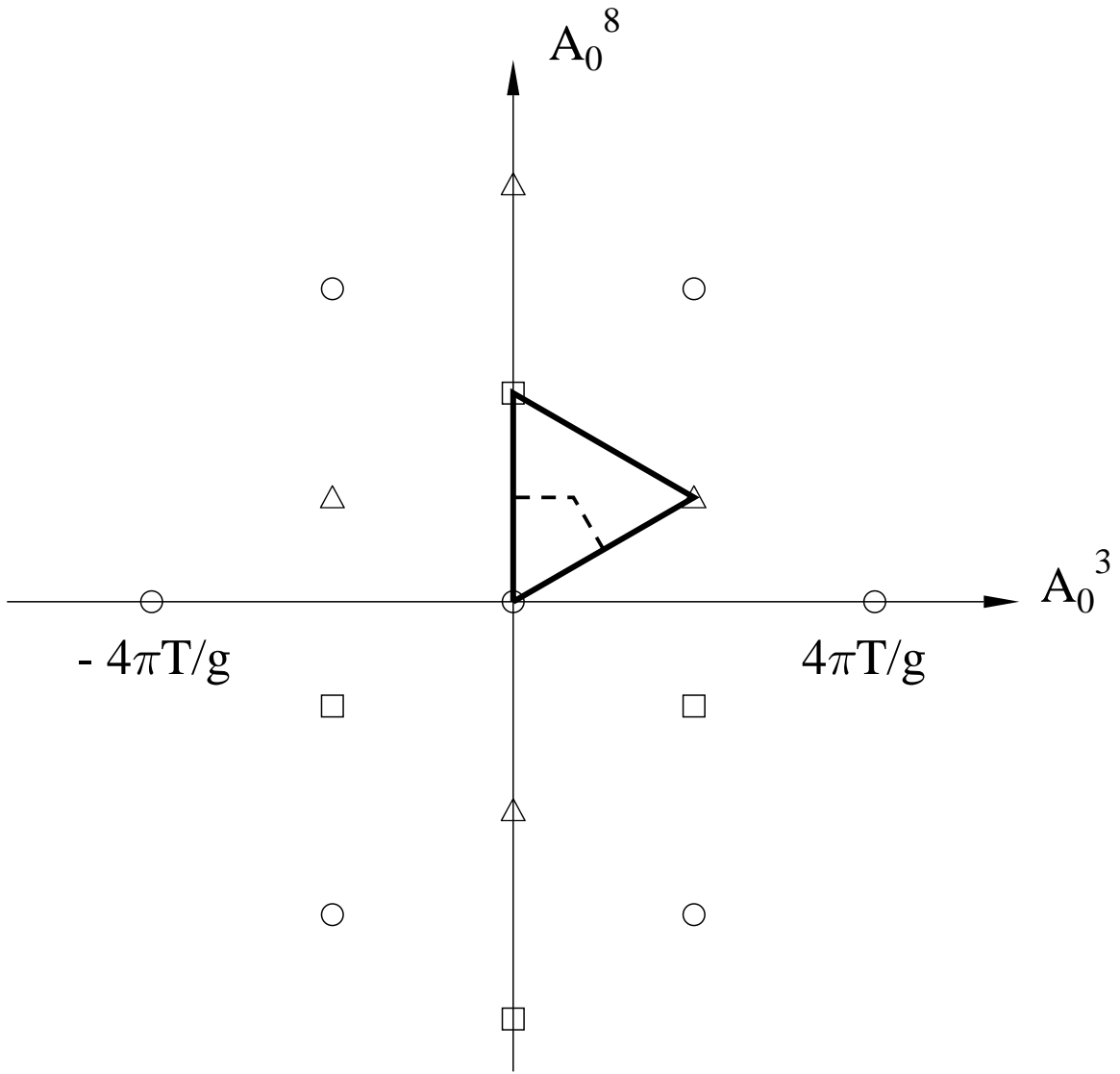
{\bf Fig.3} Geometry of effective potential for high-$T$ ~$QCD_2^{adj}$
with $N=3$.\\
 The minima occur at the points 0, $ \Box$, and $\triangle$
which are related to each other by $Z_3$ transformations and are
physically undistinguishable
in the adjoint theory. The solid triangle marks out the  standard
"fundamental" Weyl cell
and the dashed lines inside~---~ the "adjoint" Weyl cell.


\begin{thebibliography}{99}
\bibliographystyle{unsrt}

  \bibitem{Kleb} S.Dalley and I.R.Klebanov, Phys.Rev. {\bf D47}, 2517 (1993);
G.Bhanot, K.Demeterfi and I.R.Klebanov, {\bf D48}, 4980 (1993).
  \bibitem{funsp} G. 't Hooft, Nucl.Phys. {\bf B75}, 461 (1974).
  \bibitem{Hag} R.Hagedorn, Nuovo Cimento Suppl. {\bf 3}, 147 (1965);
R.Hagedorn and J.Ranft, Nuovo Cimento Suppl. {\bf 6}, 169 (1968).
  \bibitem{Yung} V.V.Khose and A.V.Yung, Z.Phys. {\bf C50}, 155 (1990).
  \bibitem{hoso} J.E.Hetrick and Y.Hosotani, Phys.Rev. {\bf D38}, 2621 (1988).
  \bibitem{Wipf}I.Sachs and A.Wipf, Helv.Phys.Acta {\bf 65}, 652(1992).
  \bibitem{cond} A.V.Smilga, Phys.Lett. {\bf B278}, 371 (1992).
  \bibitem{indus} C.Jayewardena, Helv.Phys.Acta {\bf 61}, 636 (1988).
  \bibitem{inst} A.V.Smilga, Santa Barbara preprint NSF-ITP-93-151,
 hep-th/9312110, to be published in Phys. Rev. D.
  \bibitem{flux} G.'t Hooft, Nucl.Phys. {\bf B153}, 141 (1979);
     Acta Physica Austriaca Suppl. {\bf 22}, 53 (1980).
  \bibitem{bub} A.V.Smilga, Bern preprint BUTP-93/3; Santa Barbara
     preprint NSF-ITP-93-120, {\em to be published in Ann.Phys.}.
  \bibitem{pisa} T.Bhattacharaya et al., Phys.Rev.Lett. {\bf 66}, 998(1991);
     Nucl.Phys. {\bf B383}, 497 (1992).
  \bibitem{shur} D.I.Diakonov and V.V.Petrov, Nucl.Phys. {\bf B245}, 259
(1984);
E.V.Shuryak, Rev.Mod.Phys. {\bf 65}, 1 (1993);
E.V.Shuryak and J.J.M. Verbaarschot, Nucl.Phys. {\bf B410}, 37; 55 (1993).
  \bibitem{LS} H.Leutwyler and A.V.Smilga, Phys.Rev. {\bf D46}, 5607 (1992).
  \bibitem{Coleman} S.Coleman, Commun.Math.Phys. {\bf 31}, 259 (1973).
  \bibitem{kut} D.Kutasov, Chicago preprint EFI-93-30, hep-th/9306013.
  \bibitem{kog} I.I.Kogan, Princeton preprint PUTP-1415,
           hep-th/9311164.
  \bibitem{Weiss} N.Weiss, Phys.Rev. {\bf D24}, 475 (1981); {\bf D25}, 2667
         (1982); D.J.Gross, R.D.Pisarski, and L.G.Jaffe, Rev.Mod.Phys.
          {\bf 53}, 43 (1981).
  \bibitem{Ramon} P.Ramond, {\em Field Theory: A modern primer.}
(Benjamin/Cunnings, Reading, MA, 1981).
  \bibitem{VZ} A.I.Vainshtein and V.I.Zakharov, JETP Lett., {\bf 35},
323(1982).
  \bibitem{meas} G.'t Hooft, Phys.Rev. {\bf D14}, 3432 (1976).
  \bibitem{col1} S.Coleman, {\em The uses of instantons}. In {\em Whys of
Subnuclear Physics} , Plenum Publishing Co., NY, 1979.
  \bibitem{ABC} A.I.Vainshtein et al., Sov.Phys.Usp. {\bf 25}, 195(1982).
  \bibitem{Lee} C.N.Yang and T.D.Lee, Phys.Rev. {\bf 87}, 404 (1952).
  \bibitem{Witten} E.Witten, Commun. Math. Phys. {\bf 92}, 455 (1984).
  \bibitem{Frish} Y.Frishman and J.Sonnenschein, Phys.Repts. {\bf 223},
309 (1993), {\em and references therein}.
  \bibitem{vort} A.V.Smilga, Phys.Rev. {\bf D46}, 5598 (1992).
  \bibitem{vain} M.A.Shifman, A.I.Vainstein, and V.I.Zakharov, Nucl.Phys.
{\bf B163}, 46 (1980); {\bf B165}, 45 (1980).
  \bibitem{Brown} L.S.Brown and R.I.Nepomechie, Phys.Rev.
{\bf D35}, 3239 (1987).
   \bibitem{sinst} V.Novikov et al., Nucl.Phys. {\bf B229}, 407 (1983);
G.Rossi and G.Veneziano, Phys.Lett. {\bf 138B}, 195 (1984);
D.Amati et al., Phys.Repts. {\bf 162}, 169 (1988).
   \bibitem{toron} G. 't Hooft, Comm.Math.Phys. {\bf 81}, 267 (1981).
   \bibitem{Cohen} E.Cohen and C.Gomez, Phys.Rev.Lett. {\bf 52}, 237 (1984).
   \bibitem{Zhit} A.R.Zhitnitsky, Nucl.Phys. {\bf B340}, 56 (1990).
   \bibitem{O7} M.A.Shifman and A.I.Vainstein, Nucl.Phys. {\bf B296}, 445
(1988).
   \bibitem{crew} R.J.Crewther, Phys.Lett. {\bf B93}, 75 (1980).

\end{thebibliography}
\end{document}